\documentclass[a4paper]{mem}
\usepackage{natbib}
\usepackage{graphicx}
\usepackage[a4paper]{hyperref}
\idline{73}{23}
\begin{document}
   \title{The WINGS Survey: a progress report}

   \author{G. Fasano \inst{1}, 
   B. Poggianti \inst{1},
   D. Bettoni \inst{1}, 
   E. Pignatelli \inst{1},
   C. Marmo \inst{2},\\
   M. Moles \inst{3},
   P. Kj{\ae}rgaard \inst{4},
   J. Varela \inst{3},
   W. Couch \inst{5},
   A. Dressler \inst{6} 
}

   \offprints{G. Fasano}
\mail{vicolo Osservatorio 5, 35100 Padova }

\institute{
     INAF - Osservatorio Astronomico di Padova, Vicolo Osservatorio 5, 35100 Padova, Italy \email{fasano@pd.astro.it}\\
\and Dipartimento di Astronomia, Universit\'a di Padova, Vicolo Osservatorio 2, 35100 Padova, Italy\\
\and IMAFF, CSIC, C/Serrano 123, 28006 Madrid, Spain \\
\and Copenhagen University Observatory, Juliane Maries Vej 30, 2100 Copenhagen, Denmark\\
\and School of Physics, University of New South Wales, Sydney 2052, Australia\\
\and Observatories of the Carnegie Institution of Washington, 813 Santa Barbara Street, Pasadena, CA-91101, USA
             }

   \abstract{

A two-band ($B$ and $V$) wide--field imaging survey of a
complete, all--sky X--ray selected sample of 78 clusters in the redshift 
range z=0.04--0.07 is presented. The aim of this survey is to provide the
astronomical community with a complete set of homogeneous, CCD--based
surface photometry and morphological data of nearby cluster
galaxies located within 1.5 Mpc from the cluster center. The data
collection has been completed in seven observing runs at the INT and
ESO--2.2m telescopes. For each cluster, photometric data of about 2500
galaxies (down to $V\sim$23) and detailed morphological information
of about 600 galaxies (down to $V\sim$21) are obtained by using
specially designed automatic tools.

As a natural follow up of the photometric survey, we also illustrate a
long term spectroscopic program we are carrying out with the
WHT-WYFFOS and AAT-2dF multifiber spectrographs. Star formation rates and
histories, as well as metallicity estimates will be derived for about
350 galaxies per cluster from the line indices and equivalent widths
measurements, allowing us to explore the link between the spectral
properties and the morphological evolution in high- to low-density
environments, and across a wide range in cluster X-ray luminosities
and optical properties.

   \keywords{Galaxies: Clusters -- 
                Galaxies: Evolution --
                Galaxies: Structure
               }
   }
   \authorrunning{G. Fasano et al.}
   \titlerunning{The WINGS Survey}
   \maketitle
%

\section{Introduction}

Clusters of Galaxies are the largest, yet well defined, known entities
in the Universe. The identification of their properties and content
could led to use them as tracers of cosmic evolution since they can be
detected at large distances.

Over the past five years, the cluster environment has been discovered
to be the site of morphological transformations at a relatively
recent cosmological epoch. Thanks to the high spatial resolution
achieved with the Hubble Space Telescope (HST), it has been
ascertained that the morphological properties of galaxies in the cores
of rich distant clusters largely differ from those in nearby clusters:
at $z=0.4-0.5$, spirals are a factor of 2-3 more abundant and S0
galaxies are proportionally less abundant, while the fraction of
ellipticals is already as large or larger \citep{D97,smail}.

On the basis of excellent seeing images taken at the NOT and La
Silla-Danish telescopes \citep{F02}, we have 
filled in the existing gap of information between local (z$\le$0.1)
and distant (z$\sim$0.4-0.5) clusters, confirming that, as the
redshift decreases, the S0 population tends to grow at the expense of
the spiral population \citep{FP00}. This work has also
highlighted the role played by the cluster type in determining the
relative occurrence of S0 and elliptical galaxies at a given redshift:
clusters at z$\sim$0.1-0.25 have a low/(high) S0/E ratio if they
display/(lack) a strong concentration of elliptical galaxies towards
the cluster centre.

Concerning the evolution of the stellar populations in cluster
galaxies, ground-based spectroscopic surveys of the
intermediate-redshift clusters observed by HST have offered a detailed
comparison of the spectral and morphological properties, elucidating
the link between the evolution of the stellar populations and the
changes in galaxy structure \citep{D99,PA99,CA94,CA98,FA98,LA98}.
The spiral population includes most of the star-forming
galaxies, a large number of post-starburst galaxies and a sizeable
fraction of the red, passive galaxies; in contrast, the stellar
populations of the ellipticals and of (the few) bright S0 galaxies 
appear to be old and passively evolving. These observations are
consistent with the post-starburst and star-forming galaxies being
recently infallen field spirals whose star formation is truncated upon
entering the cluster and that will evolve into S0's at a later time.

A crucial ingredient for all these evolutionary studies should be the
comprehensive knowledge of the physical properties of galaxies in
nearby clusters, in order to control their local variance, prior
to draw any conclusion on cosmic evolution. When confronted with such
kind of requirement one realizes that, while a large amount of high
quality data for distant clusters is continuously gathering from HST
imaging and ground based spectroscopy, our present knowledge of the
systematic properties of galaxies in nearby clusters remains
surprisingly poor. Actually, the only reference sample available in
the nearby universe for photometry and morphology is the
'historical' one of Dressler (1980a,b), who lists positions,
visual magnitudes and morphological classifications of galaxies,
relying on photographic plates taken at Las Campanas 2.5m, Kitt Peak
4m and Palomar 1.5m telescopes. Instead, no reference sample is
available for spectroscopy. It is obvious that an adequate
photometric and spectroscopic information on nearby clusters is
missing and is crucial for studying the morphology and the stellar
populations of galaxies in a systematic way, as well as for setting
the zero-point for evolutionary studies.

\section {The WINGS photometric survey}

   \begin{figure*}
   \centering
   \resizebox{\hsize}{!}{\rotatebox[]{-90}{\includegraphics{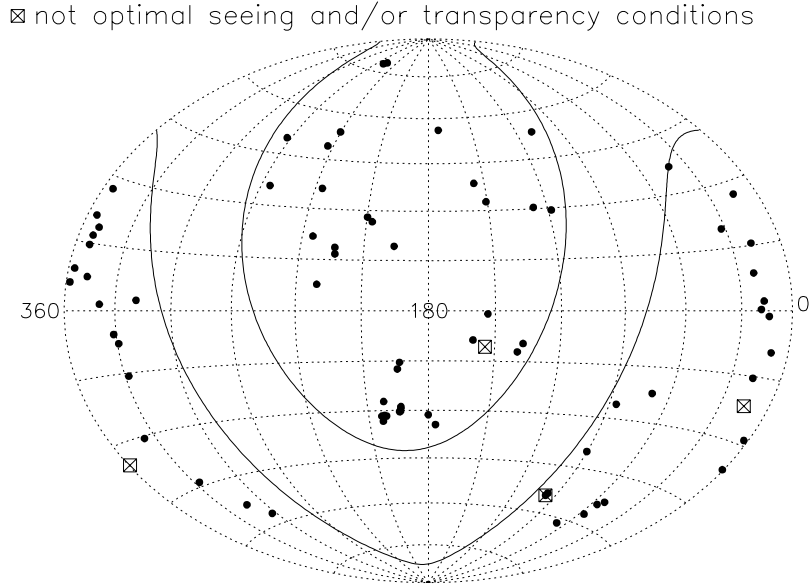}}}
   \vskip -3truecm
   \caption{All-Sky Aitoff map of the cluster sample (equatorial coordinates)}
              \label{FigAitoff}%
    \end{figure*}

The above mentioned lack of systematic information from CCD material 
on nearby galaxy clusters is mainly due to their huge angular size,
which prevented astronomers from gathering large datasets until wide
field CCD cameras (WFC) became available. In the last years excellent
WFCs became operative in imaging mode (WFI), such as those 
at the INT-2.5m and ESO-2.2m telescopes, 
as well as wide-field multifiber spectrographs, e.g.
 at the WHT-4.2m and AAT-3.9m
telescopes. We exploited the new opportunity starting with a program
named Wide--field Imaging Nearby Galaxy--cluster Survey (WINGS).
WFI proposals were presented for both the northern and the southern
hemispheres, resulting in seven obeserving runs, for a total of 18
nights (9 at the INT-2.5m telescope and 9 at the ESO-2.2m
telescope).

\subsection {The sample of nearby clusters}

   \begin{figure}
   \centering
   \includegraphics[width=6.5cm]{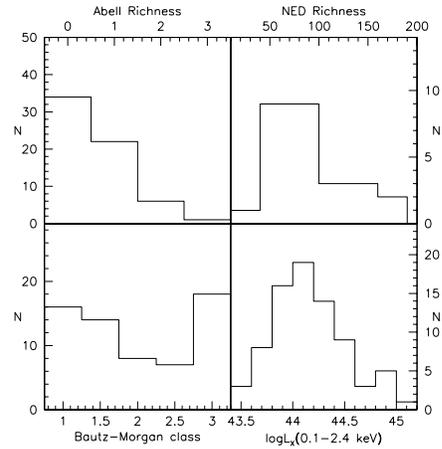}
      \caption{distribution of some cluster properties in our sample}
         \label{FigCP}
   \end{figure}

The sample we
observed has been selected from essentially complete X-ray (0.1-2.4 keV),
flux-limited samples of clusters [XBACs: Abell clusters with 
$F_X\ge$5$\times 10^{-12}$erg~cm$^{-2}$~s$^{-1}$ \citep{Eb96};
BCS+eBCS: northern clusters with 
$F_X\ge$2.8$\times 10^{-12}$erg~cm$^{-2}$~s$^{-1}$ \citep{Eb98,Eb00}] compiled
from ROSAT All-Sky Survey data. This sample is uncontaminated from
AGNs and foreground stars and is not affected by the risk of
projection effects as optically selected catalogs are.

Only clusters with galactic latitude $|b|>$20$^\circ$ in the redshift
range $0.04<z<0.07$ have been included in the sample. The redshift 
upper limit
ensures sufficient spatial resolution (1$^{\prime\prime} =$ 1.3 kpc at
$z=0.07$, $H_0$=70), while the lower limit allows us to efficiently
survey a sufficiently large area of the cluster (the central 1.5 $\rm Mpc^2$
at $z=0.04$), comparable to that observed at higher redshift with
HST. Our aim is to cover a well defined area in physical terms,
such as the $r_{200}$ radius (the radius within which the 
average cluster density is 200 times the critical density) or the virial 
radius.

The final sample includes 78 clusters (42 in the southern hemisphere
and 36 in the northern one, see Figure~\ref{FigAitoff}), of which about
a third are in common with the \citet{D80b} sample. This partial
overlap will be useful for comparing the two
data sets and cross-check the respective morphological
classifications.

Figure~\ref{FigCP} presents the distribution of some ``popular'' cluster
properties in our sample, showing that it spans a broad range in
both optical richness and X-Ray luminosity.

\subsection{Observational strategy and data reduction}

   \begin{figure*}
   \centering
   \resizebox{\hsize}{!}{\includegraphics{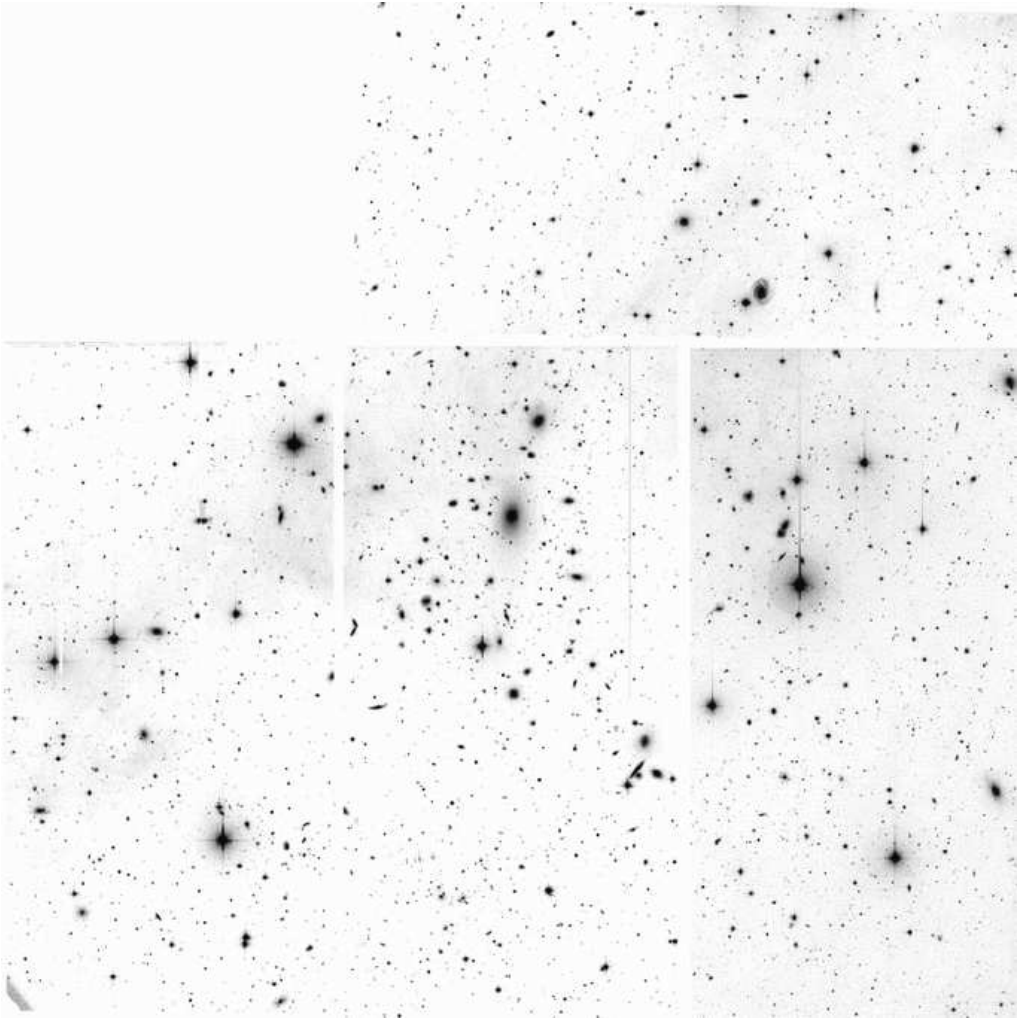}}
   \vskip 0.5truecm
   \caption{INT mosaic of the cluster Abell~2457 in the B band}
              \label{FigClu}%
    \end{figure*}

We decided to take images in the V and B bands. The former one allows
us to compare our results with previous studies of nearby clusters, as
well as with HST-WFPC2 ($I_{814}$) studies of clusters at
z$\sim$0.5. The second filter is valuable in order to get colors of galaxies
and to compare with future HST(ACS)/NGST studies of clusters at z$\sim$1.

For each cluster, three slightly offseted frames per filter have been
obtained in order to improve the cosmetics of the final mosaics. The
exposures taken with insufficient transparency and/or seeing were
repeated in different nights and/or runs, until matching the required
conditions (FWHM$\le$1.2 arcseconds). A total exposure time of 1200s
(in each band) was usually enough to reach a S/N per pixel of
$\sim$2.1(1.7) in the V(B) band. 

During each night several photometric standard fields were taken at
different positions and zenith distances in order to secure a careful
determination of the calibration coefficients as a function of
airmass, color and chip number. Also a number of astrometric and
empty fields have been observed. The latter ones will allow us to
estimate the field contamination to galaxy counts for different
bins of magnitude. The former ones have been used to determine, for
each filter, the astrometric solutions relative to each observing run. 
This is well known to be an important ingredient in the reduction 
procedure of WFI data, even as far as the photometric accuracy is
concerned. 

The standard IRAF-MSCRED package has been used for the basic reduction
procedures (debiasing, flat-fielding), while the package WFPRED \citep{RH}
was used to perform astrometry, co-adding, mosaicing
and alignment of images in different bands. A number of additional
IRAF/shell scripts have been produced in order to make 
the whole reduction procedure fully automatic. 
As an example of the final result,
Figure~\ref{FigClu} shows the INT-mosaic of the cluster Abell~2457 in
the B band.

\subsection {The Catalogs}

   \begin{figure}
   \hskip -1cm
   \includegraphics[width=7.5cm]{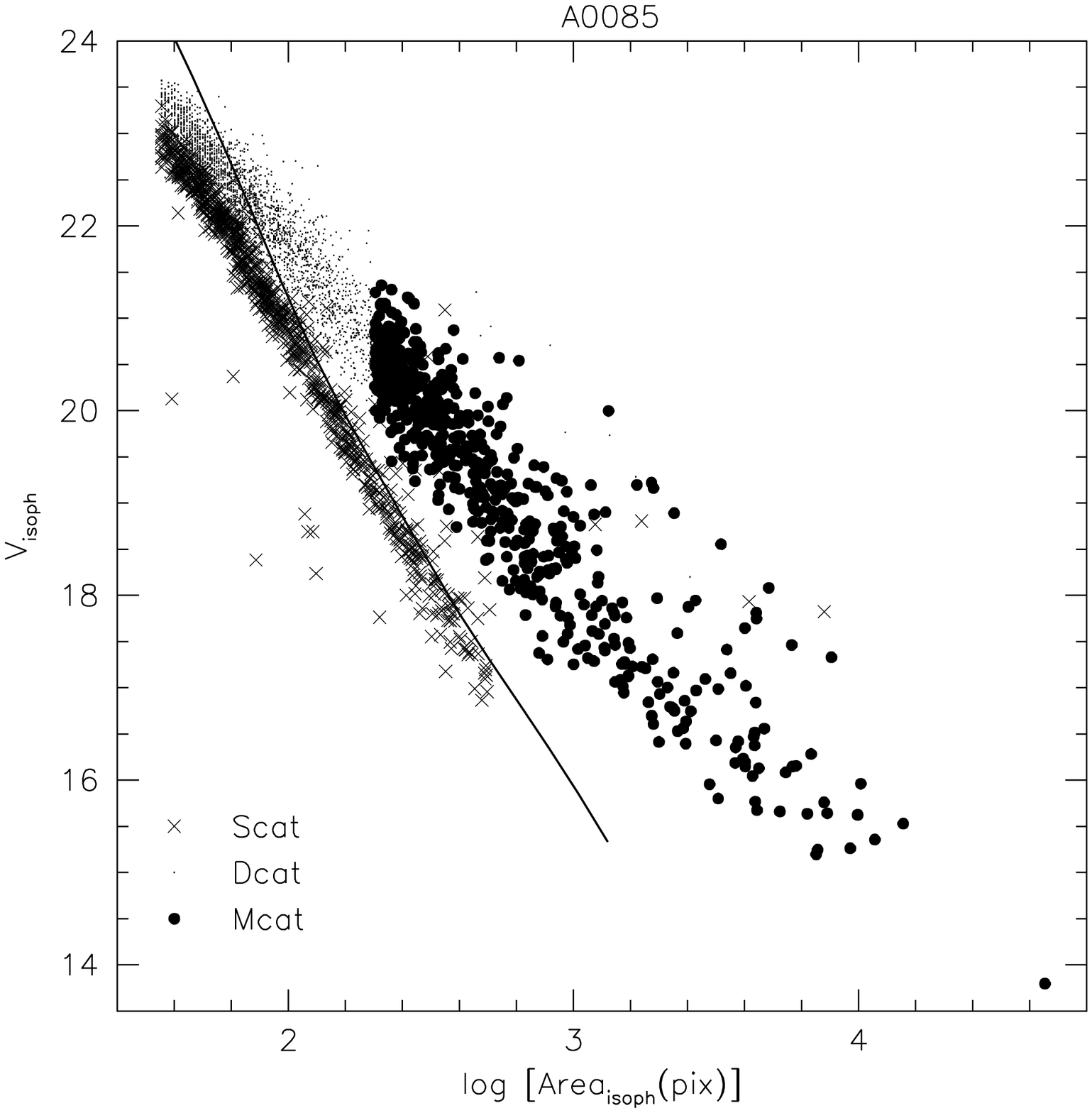}
      \caption{Automatic S/G classification compared with the expected S/G 
            partition line (see text) for the cluster Abell~85}
         \label{FigPsf}
   \end{figure}
%

   \begin{figure}
   \hskip -2cm
   \includegraphics[width=10cm]{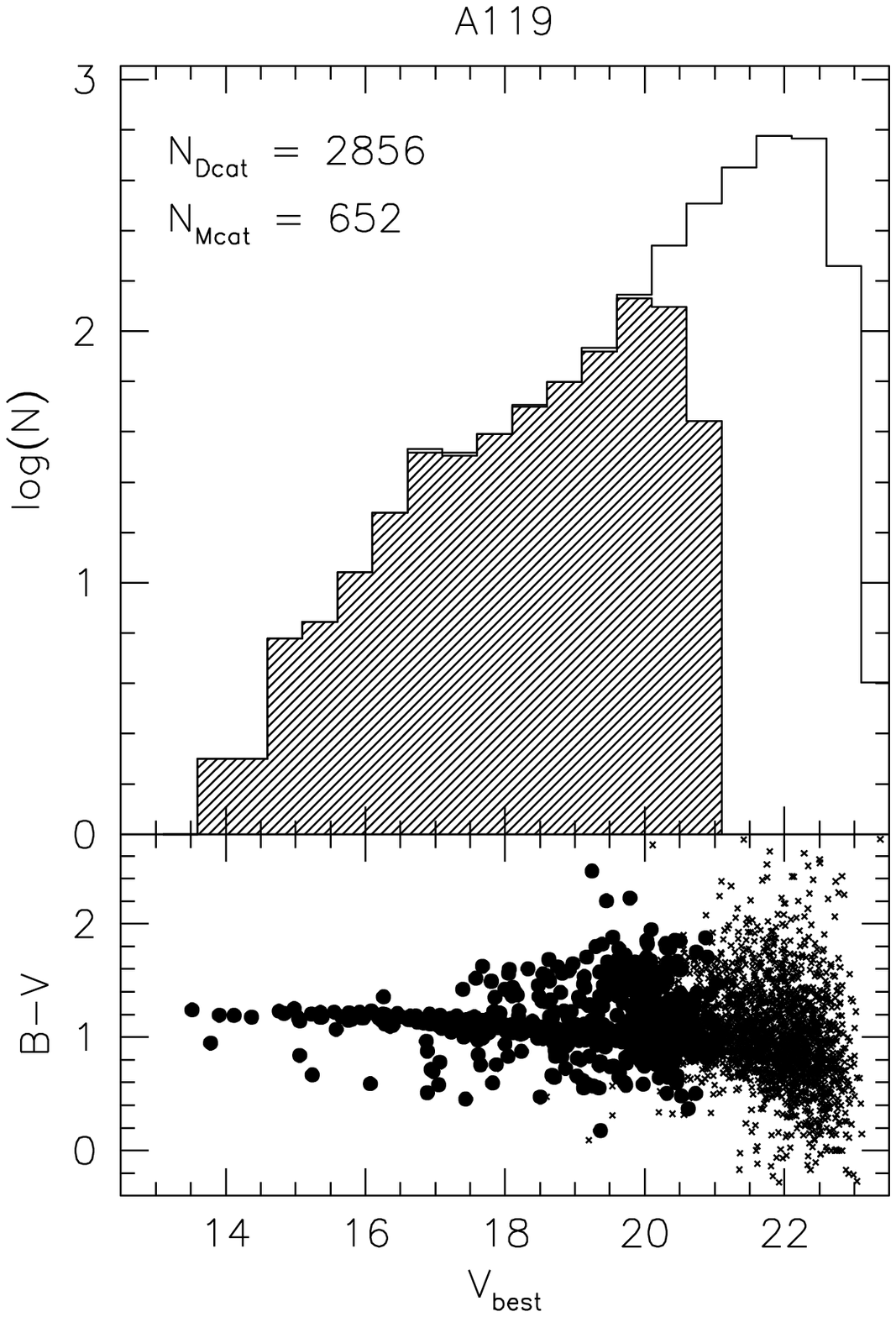}
   \vskip -0.5cm
      \caption{Histogram of magnitudes (upper panel) and color--magnitude
diagram (lower panel) of the cluster Abell~119}
         \label{FigHCM}
   \end{figure}

We used SExtractor \citep{BA96}(Bertin and Arnouts 1996) to produce a number of
source catalogs for each cluster. Relying on extensive simulations of
cluster fields, the S/N of our final V~band images turned out to be
sufficient to make SExtractor able to provide a robust
star/galaxy(S/G) separation down to a threshold (1.5$\times\sigma$)
magnitude~(area) of $\sim$23.5--24.3~(10--35~pixels), depending on
both the sky surface brightness and seeing. Thus, for each cluster
mosaic in the V band, the proper extraction limits of magnitude and
area have been derived and four deep catalogs have been
produced for: galaxies (Dcat; S/G$\le$0.2), stars (Scat; S/G$\ge$0.8),
other objects (Ocat; 0.2$<$S/G$<$0.8) and saturated sources
(Satcat). Almost all objects with V$>$23 turned out to belong to the
Ocat catalog. An additional catalog for surface photometry and morphology of
galaxies (Mcat; see next subsection) has been extracted from Dcat
including only galaxies with threshold area~A$\ge$200
pixels. Besides the different kinds of magnitudes (aperture,
isophotal, total), all derived neglecting the color term, these catalogs
contain some global information about size, ellipticity and position
angle of objects. The V band catalogs have been then used as reference
lists to extract the corresponding catalogs in the B band, allowing us
to measure the 5-kpc ($H_0$=70) aperture colors and to evaluate the
color corrected magnitudes in both the V and B bands.

The automatic S/G classifier can be sometime worsened by
space--varying PSFs and blending. It has been improved interactively
looking at the position in a plane like that of Figure~\ref{FigPsf}
(threshold area--V$_{mag}$), where the Star/Galaxy partition line,
analitically derived from the proper PSF, is reported for the cluster
Abell~85. This interactive cleaning task moves some automatically
classified galaxies in the star catalog and viceversa.

In the upper panel of Figure~\ref{FigHCM} we report the distributions
of the total SExtractor magnitudes (V$_{best}$) from the galaxy
catalogs Dcat and Mcat, relative to the cluster Abell~119. In the same
panel also the number of galaxies in the two
catalogs are indicated. The lower panel of the same figure illustrates
the corresponding color--magnitude plot. Figure~\ref{FigHCM} shows
that the completeness of the bright and deep galaxy catalogs (Mcat and
Dcat) is typically achieved down to V$\sim$20 and V$\sim$22,
respectively, while the corresponding cutoff magnitudes turn out to be
tipically $\sim$1~mag fainter. Also typical are the sizes of the
catalogs Mcat and Dcat indicated in the figure.  

It is worth to note that, due to the extrapolation of the luminosity
profiles (assumed gaussian), the total galactic magnitudes given by
SExtractor could be over--estimated up to 0.4-0.5~mag, the
early--types being more biased than late--types \citep{F98}.
This bias disappears if we consider the magnitudes from our
surface photometry tool (see next subsection). However,
the magnitudes of galaxies not belonging to the catalog Mcat need to be
statistically corrected for the bias using concentration indices.
\subsection{Surface photometry and morphology}

   \begin{figure}
   \includegraphics[width=6.5cm]{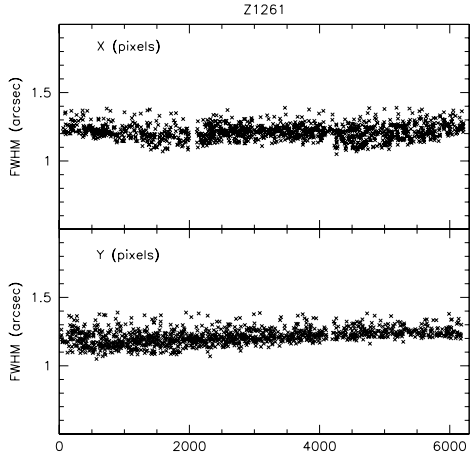}
      \caption{FWHM along the X and Y axes for the cluster Z1261}
         \label{FigFwhm}
   \end{figure}

The catalog Mcat contains those galaxies from the deep list (Dcat) that
are large
enough (A$\ge$200~pixels) to be suitable for surface photometry and
morphological analysis. This defines the reference list of galaxies to
be processed by the automatic surface photometry tool GASPHOT.

\begin{figure}
   \vskip -2cm
   \hskip -2cm
   \includegraphics[width=10cm]{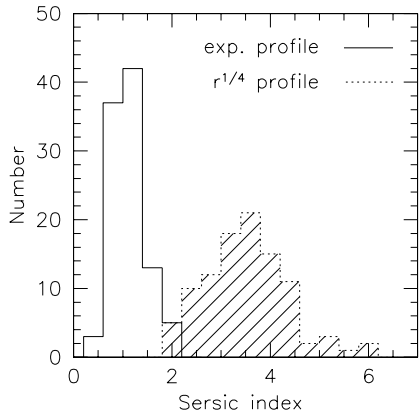}
   \vskip -2cm
      \caption{Histograms of the Sersic index obtained by running
      GALPROF on a crowded artificial field containing 200 toy
      galaxies with different magnitudes and radii. Galaxies with
      exponential profiles (100 objects) are binned with the solid
      line histogram, whereas the dotted line histogram refers to
      galaxies with r$^{1/4}$ profiles.}
         \label{FigSers2}
   \end{figure}

The need for such a software has become more and more evident
in the last years, as deep and/or wide imaging became more and
more common. The usual 'one at a time' surface photometry tools
(IRAF-ELLIPSE, Jedrzejewski 1987; AIAP, Fasano 1990) are clearly
inadequate to process CCD frames containing several hundreds (or even
thousands) of galaxies. For this reason we have developed a Galaxy
Automatic Surface PHOtometry Tool (GASPHOT, Pignatelli \& Fasano 1999)
which is able to process 'all at once' a galaxy list like Mcat.
It consists of four main tools:

\begin{itemize}

\item STARPROF produces a space--varying PSF profile. This
is achieved by using an analytical (multi--gaussian) representation of
the PSF and turns out to be crucial in order to obtain reliable
results in the morphological analysis (next steps), particularly when
the variation of the FWHM along the axes is not negligible (see
Figure~\ref{FigFwhm});

\item SEXISOPH exploits  SExtractor capabilities to
produce luminosity, ellipticity and position angle profiles of the
whole galaxy sample;

   \begin{figure*}
   \centering
   \resizebox{\hsize}{!}{\rotatebox[]{-90}{\includegraphics{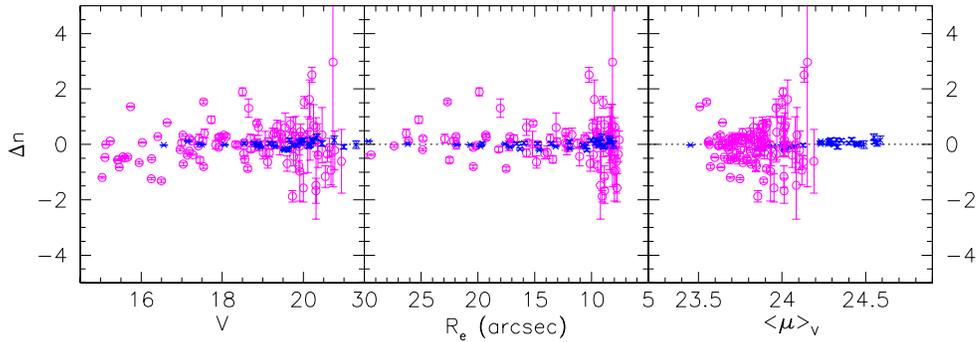}}}
   \vskip -6.5truecm
   \caption{Sersic index errors resulting from a GALPROF run on
   a grid of 200 toy galaxies with different magnitudes and radii.
   Exponential profiles (starred points) are much more precisely 
   recovered with respect to the r$^{1/4}$ profiles (open circles).}
              \label{FigSers1}%
    \end{figure*}

\item GALPROF simultaneously fits the major and minor axis 
luminosity profiles of each galaxy by using a Sersic $r^{1/n}$ law
and/or a two component ($r^{1/4}$ + exponential) profile, convolved in
any case with the proper PSF. At variance with SExtractor, GALPROF
produces unbiased estimates of the total magnitudes, independently on
the morphological type. Figures~\ref{FigSers1} and \ref{FigSers2}
illustrate the performances of GALPROF in recovering the Sersic's
index of toy galaxies with r$^{1/4}$ and exponential luminosity
profiles. 

\item MORPHOT exploits some characteristic 
features of the luminosity and geometrical profiles, together with the
Sersic index, to estimate the morphological type of individual
galaxies.

\end{itemize}

For each cluster GASPHOT produces two more catalogs. The first one
includes the whole set of luminosity and geometrical profiles obtained
by SEXISOPH. The second one contains the photometric and morphological
information extracted from the profiles by GALPROF and
MORPHOT (effective radius and average surface brightness, total
magnitude, Sersic index, B/D ratio, morphological type, etc..).

\begin{figure}
   \includegraphics[width=6.5cm]{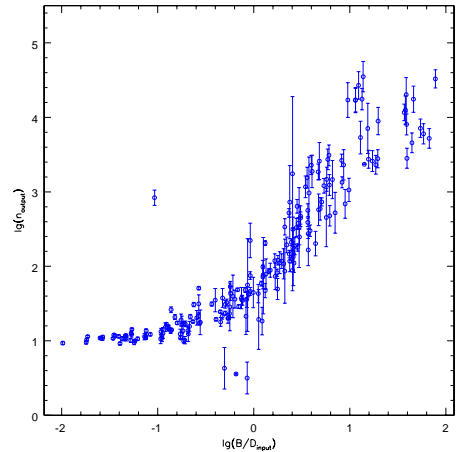}
      \caption{Correlation between the Sersic index $n$ resulting from
      GALPROF and the 'true' Bulge/Disk ratio for a sample of faint
      toy galaxies}
         \label{FigSers}
   \end{figure}

Luminosity profiles of faint and/or small galaxies are usually well
enough represented by a simple Sersic law, even in case of
multi--component objects (see Figure~\ref{FigSers}). On the contrary,
for bright and/or big galaxies two components are often necessary to
model the luminosity profiles.

%

\subsection {Status and perspectives}

   \begin{figure*}
   \centering
   \resizebox{\hsize}{!}{\includegraphics{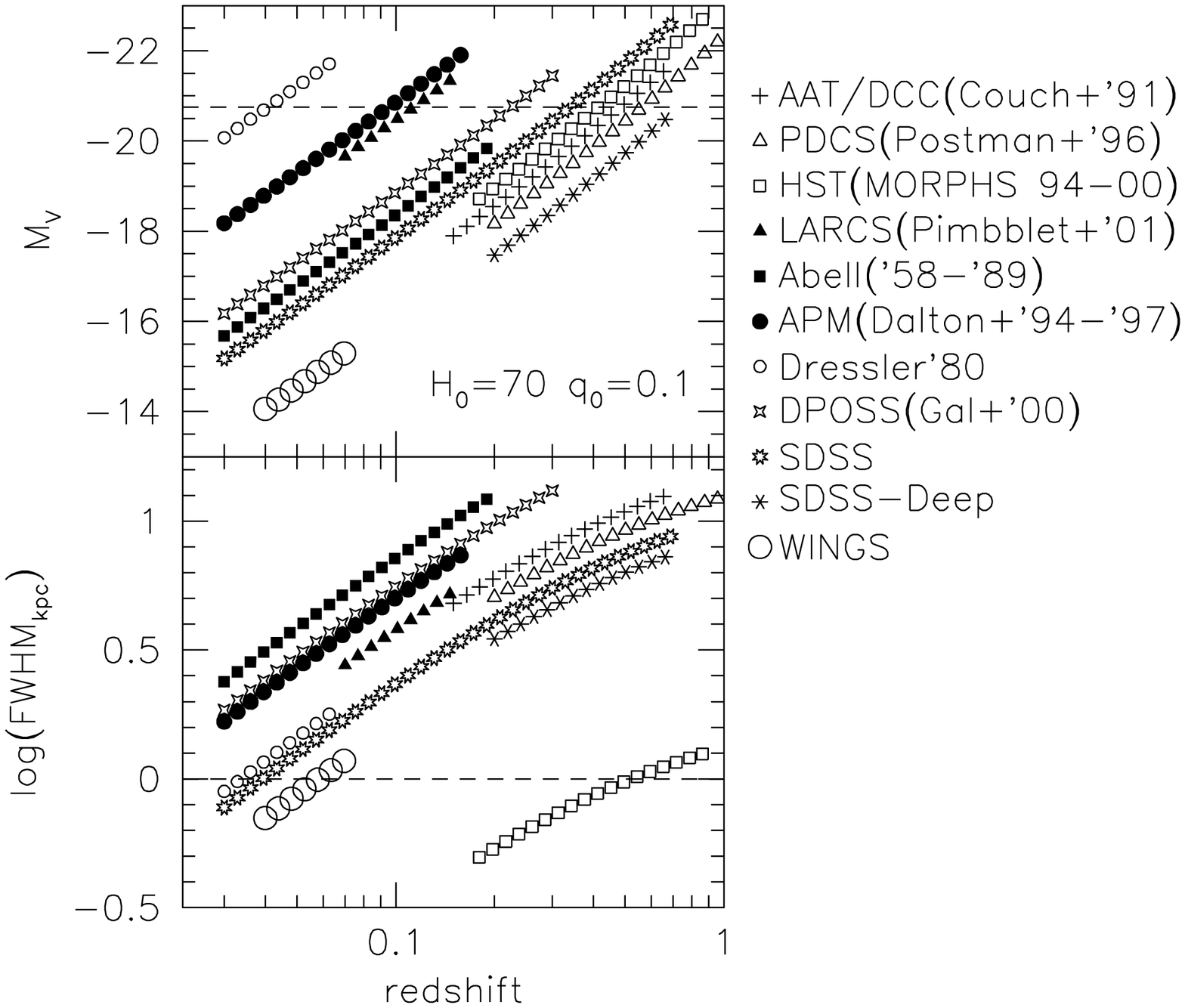}}
   \vskip -0.5truecm
   \caption {Limiting absolute magnitude and space resolution (in
   Kpc) as a function of redshift for most of the available and
   ongoing extragalactic surveys.}
              \label{FigWings}%
    \end{figure*}

The data collection for the photometric survey has been completed
in August 2002. The vast majority of the clusters in the sample 
($>$90\%) have been observed in good (sometime very good) weather
conditions, making us confident that reliable photometric and 
morphological results will be obtained. 

The INT--WFI frames have been fully reduced and most of the
corresponding catalogs have been already extracted. They will be
included in the first paper of the WINGS series (Fasano et al. 2003, in
preparation). The morphological
analysis of these clusters, which is presently being carried out with
GASPHOT, will be the subject of a forthcoming paper. Concerning the
ESO-WFI frames, the data reduction is currently in progress.

As stated in Section~1, The photometric WINGS survey has been
conceived to fill in the lack of a systematic investigation of nearby
clusters and their galaxy content. This is schematically illustrated
in Figure~\ref{FigWings}, where the limiting absolute magnitude and
space resolution (in kpc) are reported as a function of redshift for
most of the available and ongoing galaxy surveys. It can be
seen that WINGS is the deepest (M$_V\sim$-14), best resolution
(FWHM$\sim$1~kpc) survey of a complete sample of galaxies in nearby
clusters to date. For instance, even if the nominal resolution
(FWHM in kpc) of WINGS is only slightly better than that of the
survey of Dressler (1980), its data quality (CCD) is
definitively better and its
deepness is incomparably better ($\sim$6~mag) with respect to the
Dressler's survey. 

This survey will produce detailed surface photometry and morphology
for about 5$\times$10$^4$ galaxies, while integrated photometry and
rough structural parameters will be obtained for about
2$\times$10$^5$ galaxies. This will represent the first CCD-based
systematic investigation of the properties of cluster galaxies in the
nearby universe and will provide a local reference sample for distant
cluster studies.

What could we do with this huge dataset~? The only limit we have is
our imagination~!
 
In the next few years deep imaging surveys of distant clusters will
become an even stronger scientific drive for HST with the new Advanced
Camera and will be paralleled by extensive spectroscopic surveys with
8-m class telescopes. While the high-z studies occupy the headlines,
the knowledge of the local universe is crucial to allow a full
exploitation of the high-z data. Given the large cluster-to-cluster
variations in properties and morphological content, a large and
well-defined sample is needed to investigate in a systematic way what
cluster property/ies are driving the variations in galactic
properties. 

We can do that~!
In fact, apart from a robust analysis of the global cluster properties
(characteristic radius, total luminosity, ellipticity and shape,
subclustering, etc..), this dataset will allow us to study with high
statistical significance the global properties of cluster galaxies:
luminosity function of different morphological types, morphological
fractions E:S0:Sp:Irr, integrated colors, colour-magnitude diagrams,
scaling relations ($<\mu_e>-r_e$--Sersic~index).
Besides, we will be able to look at all the above mentioned galactic
properties as a function of the cluster properties (structure and
concentration, X-ray and total optical luminosity), local density and
clustercentric distance.

We plan to release the WINGS catalogs to the astronomical community
in a couple of years.

\section {The WINGS spectroscopic survey}
The spectroscopic follow-up of WINGS (WINGS-SPE) is
obtaining optical spectra of 300 to 500 galaxies per cluster in a
statistically significant subsample of the WINGS clusters.  This
subsample consists of about 55 clusters with a range of X-ray and
optical properties sufficiently large to explore the dependence on
cluster properties. Clusters are known to lie on a ``fundamental
plane'' in a three-dimensional parameter space identified by optical
luminosity, half-light radius and X-ray luminosity (e.g. Miller et
al. 1999, Fritsch \& Buchert 1999), and the deviations from such a
plane are strongly correlated with substructure in the
cluster. The cluster selection for WINGS-SPE ensures a coverage of
this fundamental-plane over a factor of 30 in X-ray luminosity.

The only criterion for spectroscopic target selection is the galaxy
total magnitude limit $V<20$, corresponding to $M_B \sim -17$ (on average)
over our redshift range. In addition, galaxies lying above the
color-magnitude sequence in the B-V color magnitude diagram are
sampled at a lower completeness rate than those on and below the sequence.
This selection criteria provide an unbiased magnitude-limit
sample of galaxies representative of the whole cluster populations,
while enhancing the probability to reject non-cluster members.

The magnitude limit for spectroscopy reaches more than 5 magnitudes 
down the galaxy luminosity function, thus 1.5 magnitude deeper
than large area spectroscopic surveys such as the Sloan or the 2dF
Galaxy Redshift Survey. The depth of the spectroscopy is important
for two reasons. First of all, because it allows an unprecedented
view of both massive and intermediate-mass galaxies in clusters,
allowing an investigation of the stellar content and morphology
as a function of the galaxy luminosity. Second, a wide magnitude
coverage is needed for a useful comparison with distant cluster
studies. In fact, a large fraction of the luminous star-forming
galaxies at high z are expected to fade significantly as a consequence of
the decline in their star formation, thus populating
the intermediate-to-faint luminosity regime in the nearby clusters.

WINGS-SPE is a long term spectroscopic campaign that has begun in 
semester 2002B and will stretch over (at least) two
semesters. Fifteen nights of spectroscopy have been allocated so far.
The spectra are obtained with a multifiber technique with the WYFFOS
spectrograph at the William Herschel Telescope and the 2dF
spectrographs at the Anglo-Australian Telescope. The use of both
facilities is vital for the project, because the Northern and the
Southern emispheres contain the X-ray faintest and X-ray brightest
subsets of the clusters, respectively.

Spectra cover the range 3600-8000 \AA $\,$ (with 2dF) and 
3800-7000 \AA $\,$ (with WYFFOS), at a resolution of 9 and 6 \AA,
respectively. The wide magnitude range of the galaxies requires
two different fibre configurations per cluster with different exposure times.
From the spectra we are measuring redshifts, line indices and equivalent 
widths of the main absorption and emission features, which provide
cluster membership, star formation rates and histories, and
metallicity estimates.

Obviously, this dataset is suitable for addressing numerous scientific issues.
Our primary goal is to study the issues described below.

\subsection{The link between star formation and galaxy morphology in
dense environments.}  
The WINGS sample is unique in terms of the
detailed, quantitative morphological information that is available for
its galaxy populations. This virtue is particularly important when
combined with similarly detailed information about the star formation
activity and metal content of these galaxies.
 
Extensive work on galaxy clusters at high redshift has demonstrated that
the combination of morphological and spectroscopic information is a powerful 
tool in the study of galaxy evolution (Dressler et al. 1999; Poggianti et 
al. 1999; Couch et al. 1998). High-$z$ observations suggest that stellar 
populations in cluster ellipticals are old, while star formation in disk 
galaxies had a steeper evolution in clusters than in the field.

A similar study is missing in the nearby Universe. 
Studies of a few low-$z$ clusters (Kuntschner \& Davies 1998,  
Smail et al. 2001, Poggianti et al. 2001) have begun to uncover 
a difference in the age distributions of S0 and E galaxies, which is 
consistent with the hypothesis that star-forming spirals are transformed
into passive S0s. 

Based on WINGS spectra, we are currently exploring the stellar ages
and metallicity of galaxies as a function of their Hubble type and 
luminosity. This will clarify whether the 
differences between S0s vs Es are a widespread phenomenon in clusters,
will elucidate the spectroscopic properties of the cluster
spirals and what is the incidence of passive, 
red spirals in high- to low-density regions.

\subsection{The dependence of star formation on clustercentric
position, local density and cluster properties.}  
The effects that
different physical mechanisms (ram pressure stripping, tidal
encounters, loss of halo gas etc.) have on galaxy evolution are
expected to vary with the local density and/or the global cluster
properties and accretion history. Having a wide-area dataset of a
large and variegate sample of clusters, we hope to isolate the
different effects to understand what processes have a noticeable
influence on galaxy evolution, and how they modify galaxies' star
formation history. In fact, we are carrying out a simultaneous analysis
of the galactic properties {\bf and} their environment, with the aim of
understanding how galaxy evolution is related with the cluster/group
mass, with the amount of substructure and ongoing merging of groups and
clusters, with the intracluster medium local density and metallicity.

\subsection{The comparison with distant clusters.} 
Finally, WINGS-SPE represents the first local spectroscopic
database of its kind
that can be used as a baseline for comparison with clusters at higher 
redshift. For this work, aperture effects need to be taken into account,
since slit spectra at high z typically cover much
larger galactic areas than the fibre spectra at low z. However, this 
problem is mitigated by the fact that 
aperture effects can be estimated from radial color gradients within
each galaxy, based on the precision photometry of WINGS, using
the fact that color and spectral type are largely correlated.

\begin{acknowledgements}
      Part of this work was supported by the Italian Scientific Research
      Ministery (MIUR).
\end{acknowledgements}

\bibliographystyle{aa}

\begin{thebibliography}{}

\bibitem[Bertin \& Arnouts (1996)]{BA96}
Bertin E., Arnouts S. 1996, A\&AS 117, 393

\bibitem[Couch et al. (1994)]{CA94}
Couch, W.J., Ellis, R.S., Sharples, R.M., Smail, I., 1994, ApJ, 430, 121 

\bibitem[Couch et al. (1998)]{CA98}
Couch, W.J., Barger, A.J., Smail, I., Ellis, R.S., Sharples, R.M. 1998, ApJ 497, 188 

\bibitem[Dressler (1980)]{D80a} 
Dressler, A. 1980a, ApJ 236, 351

\bibitem[Dressler (1980)]{D80b} 
Dressler, A. 1980b, ApJS 424, 565

\bibitem[Dressler et al. (1997)]{D97} 
Dressler, A., Oemler, A.Jr., Couch, W.J., Smail, I., Ellis, R.S., 
Barger, A., Butcher, H., Poggianti, B.M., Sharples, R.M. 1997, ApJ 490, 577

\bibitem[Dressler et al. (1999)]{D99} 
Dressler, A., Smail, I., Poggianti, B.M., Butcher, H., Couch, W.J.,
Ellis, R.S., Oemler, A.Jr. 1999, ApJS 122, 51 

\bibitem[Ebeling et al. (1996)]{Eb96} 
Ebeling, H., Voges, W., Bohringer, H., Edge, A.C., Huchra, J.P., Briel, U.G. 1996, MNRAS 281, 799

\bibitem[Ebeling et al. (1998)]{Eb98} 
Ebeling, H., Edge, A.C., Bohringer, H., Allen, S.W., Crawford, C.S., Fabian, A.C., 
Voges, W., Huchra, J.P. 1998 MNRAS 301, 881

\bibitem[Ebeling et al. (2000)]{Eb00} 
Ebeling, H., Edge, A.C., Allen, S.W., Crawford, C.S., Fabian, A.C., Huchra, J.P. 2000, MNRAS 318, 333

\bibitem[Fasano (1990)]{F90} 
Fasano, G. 1990, Internal report of the Padova Astronomical Observatory

\bibitem[Fasano et al. (1998)]{F98} 
Fasano, G., Cristiani, S., Arnouts, S., Filippi, M. 1998, AJ 115, 1400

\bibitem[Fasano et al. (2000)]{FP00} 
Fasano, G., Poggianti, B.M., Couch, W.J., Bettoni, D., Kj{\ae}rgaard, P., Moles, M. 2000, ApJ 542, 673

\bibitem[Fasano et al. (2002)]{F02} 
Fasano, G., Bettoni, D., D'Onofrio, M., Kj{\ae}rgaard, P., Moles, M. 2002, A\&A 387, 26

\bibitem[Fisher et al. (1998)]{FA98} 
Fisher, D., Fabricant, D., Franx, M., van~Dokkum, P. 1998, ApJ 498, 195

\bibitem[Fritsch \& Buchert (1999)]{fb99}
Fritsch \& Buchert 1999, A\&A, 344, 749

\bibitem[Jedrzejewski (1987)]{Jed} 
Jedrzejewski, R. 1987, MNRAS 226, 747

\bibitem[Miller et al. (1999)]{miller99}
Miller, Melott \& Gorman, 1999, ApJL, 526, L61

\bibitem[Lubin et al. (1998)]{LA98} 
Lubin, L.M., Postman, M., Oke, J.B., Ratnatunga, K.U., Gunn, J.E., Hoessel, J.G., Schneider, D.P. 1998, AJ 116, 584

\bibitem[Pignatelli \& Fasano (1999)]{Pign} 
Pignatelli, E., Fasano, G. 1999, Ap\&SS 269, 657

\bibitem[Poggianti et al. (1999)]{PA99} 
Poggianti, B.M., Smail, I., Dressler, A., Couch, W.J., Barger, A.J.,
Butcher, H., Ellis, R.S., Oemler, A.Jr. 1999, ApJ 518, 576 

\bibitem[Poggianti et al. (2001)]{P01}
Poggianti, B.M., Bridges, T.J., Carter, D., Mobasher, B., Doi, M., Iye, 
M., Kashikawa, N., Komiyama, Y., Okamura
, S., Sekiguchi, M., Shimasaku, K., Yagi, M., Yasuda, N., 2001, 
ApJ, 563, 118 


\bibitem[Rizzi \& Held (2002)]{RH} 
Rizzi, L., Held, E.V. 2002, in preparation

\bibitem[Smail et al. (1997)]{smail} 
Smail, I., Dressler, A., Couch, W.J., Ellis, R.S., Oemler,
A.Jr., Butcher, H., Sharples, R.M. 1997, ApJS 110, 213

\bibitem[Smail et al. (2001)]{smail01}
Smail, Kuntschner, Kodama et al. 2001, MNRAS, 323, 839

\end{thebibliography}

\end{document}